\documentclass[twocolumn,showpacs,preprintnumbers,amsmath,amssymb]{revtex4}

\usepackage{graphicx}
\usepackage{dcolumn}
\usepackage{bm}
\usepackage{color}

\begin{document}


\title{Transition between Quantum States in a Parallel-Coupled 
Double-Quantum-Dot
}

\author{J.C. Chen$^1$}
\author{A.M. Chang$^1$}%
 \email{yingshe@physics.purdue.edu}
\author{M.R. Melloch$^2$}
\affiliation{%
$^1$Department of Physics, Purdue University, West Lafayette, IN 47907\\
%
$^2$School of Electrical and Computer Engineering, Purdue University, 
West Lafayette, IN 47907\\
}%

\date{\today}

\begin{abstract}

Strong electron and spin correlations in a double-quantum-dot (DQD) can give 
rise to different quantum states.  We observe a continuous transition from a 
Kondo state exhibiting a single-peak Kondo resonance to another exhibiting a 
double peak by increasing the inter-dot-coupling (t) in a parallel-coupled DQD.  
The transition into the double-peak state provides evidence for 
spin-entanglement between the excess-electron on each dot.  Toward the 
transition, the peak splitting merges and becomes substantially smaller than 
t because of strong Coulomb effects.  Our device tunability bodes well for 
future quantum computation applications.

\end{abstract}

\maketitle


The double-quantum-dot (DQD) is emerging as a versatile system for studying 
a variety of strongly correlated behaviors \cite{1,2,3,4,5,6,7}.  Following 
the experimental demonstration of the Kondo impurity-spin screening effect 
in single quantum dots \cite{8,9,10,11,12,13}, recent theoretical investigations 
of the coupled-DQD system is uncovering new correlated behaviors 
\cite{1,2,3,4,5,6,7}.  These works suggest that the DQD enables a realization 
of the two-impurity Kondo problem first discussed in the context of metallic 
systems \cite{1,14,15,16} in which a competition between Kondo correlations 
and antiferromagnetic (AF) impurity-spin correlation leads to a quantum 
critical phenomenon.  In a different regime of parameters, a related quantum 
critical phenomenon can occur driven by a competition between intra-dot Kondo 
coupling to leads and the inter-dot-coupling \cite{2,3}.  In each scenario, 
a transition is predicted to occur between a quantum state characterized by 
a single-peaked Kondo resonance, and a different state with a double-peaked 
resonance.  Depending on model and DQD geometry--whether series or parallel 
coupled--both a continuous or discontinuous \cite{6} behavior in the Kondo 
peak characteristics have been predicted.  
The quantum transition
in the two-impurity Kondo problem has received wide attention in the 
theoretical literature in the past two decades, to a large extent because
the Kondo to antiferromagnetic transition involves an unusual
{\it non-Fermi liquid} fixed point.
Experimental investigation 
of this problem thus far has not been reported. 

     Here we describe transport properties of an artificial molecule formed 
by two-path, parallel-coupled double-quantum-dots, where the 
inter-dot-tunnel-coupling, t, can be tuned.  In the Kondo regime the 
differential-conductance, dI/dV, exhibits a single peak centered at 
zero-bias for t comparable to the lead-coupling induced level broadening.  
Increasing t by less than 10\% resulted in a continuous evolution into a 
split Kondo resonance.  At the same time the conductance at zero-bias 
exhibits a maximum in the vicinity of the transition.  This peak splitting
behavior in conjunction with distinct temperature dependences in the different
regimes demonstrates a direct observation of an inter-dot-coupling-induced
quantum transition.  Moreover, on the double peak side
{\it the zero-bias conductance becomes suppressed; this suppression
represents direct evidence that the localized dot spins are becoming
entangled into a spin singlet}.
     
     While a double-peaked, coherent Kondo effect was observed by Jeong 
{\it et al.} in a series-coupled DQD geometry\cite{17}, the existence of a 
single-peaked Kondo effect at finite inter-dot-coupling has not previously 
been established.  The series-geometry is unsuitable for an investigation of 
the transition because even a slight decrease in t can drastically reduce 
the tunnel current below detection.  Therefore neither a single-peaked behavior 
nor a transition could be observed. Furthermore, direct evidence for
spin-entanglement was not obtainable.  In the parallel geometry the
undesirable suppression of the conductance is avoided.
          
     The Kondo effect in a single QD results from the coupling between the 
dot excess (unpaired) spin and the spin of the conduction electrons in the 
leads.  The energy scale is given by the Kondo temperature $T_K\approx 
\sqrt{U\Gamma} exp[-\pi|\mu - \epsilon_0|(U + \epsilon_0)/\Gamma U]$ where 
U is the on-site charging energy, $\epsilon_0$ is the energy of the 
single-particle level, $\Gamma$ reflects the dot level broadening from 
coupling to the leads, and $\mu$ is the chemical potential. The fully 
symmetric DQD system contains the additional inter-dot-coupling parameter, t.  
The two magnetic impurities, realized by a single excess-spin on each dot, 
interact through an effective AF coupling, $J=4t^2/U$. The new energy scales, 
t and J, together with U, $\epsilon_0$, and $\Gamma$, introduce a rich variety 
of correlated physics.  When U is the dominant energy scale as in the case of 
our quantum dots, for energies below U two different scenarios are predicted 
for dots coupled in series \cite{1,2,3,4,5,6}.  When t$ < \Gamma$, the system 
can be cast \cite{1} into the two-impurity Kondo problem initially discussed 
by Jones et al. \cite{14,15,16}. The competition between Kondo effect and 
antiferromagnetism appears as a continuous phase transition (or crossover) 
at a critical value of the coupling $J/T_K \approx 2.5$. If however, $t$ is 
tuned to $t > \Gamma$  before the AF transition point can be reached, the 
system undergoes a continuous transition from a separate Kondo state of 
individual spins on each dot (atomic-like) to a coherent bonding-antibonding 
superposition of the many-body Kondo states of the dots  (molecular-like) 
\cite{2,3,7}.   Both the AF state and coherent bonding state exhibit a 
double-peaked Kondo resonance in the differential conductance versus 
source-drain bias and involve entanglement of the dot spins into a 
spin-singlet.  Therefore they are likely closely related to one another.  
The parallel-coupled case has only recently been analyzed in a model 
without inter-dot-tunnel-coupling, t, where the AF coupling occurs via 
electrostatic coupling.  A discontinuous jump in the differential 
conductance is predicted when the AF state becomes favored \cite{6}. 

\begin{figure}
\includegraphics{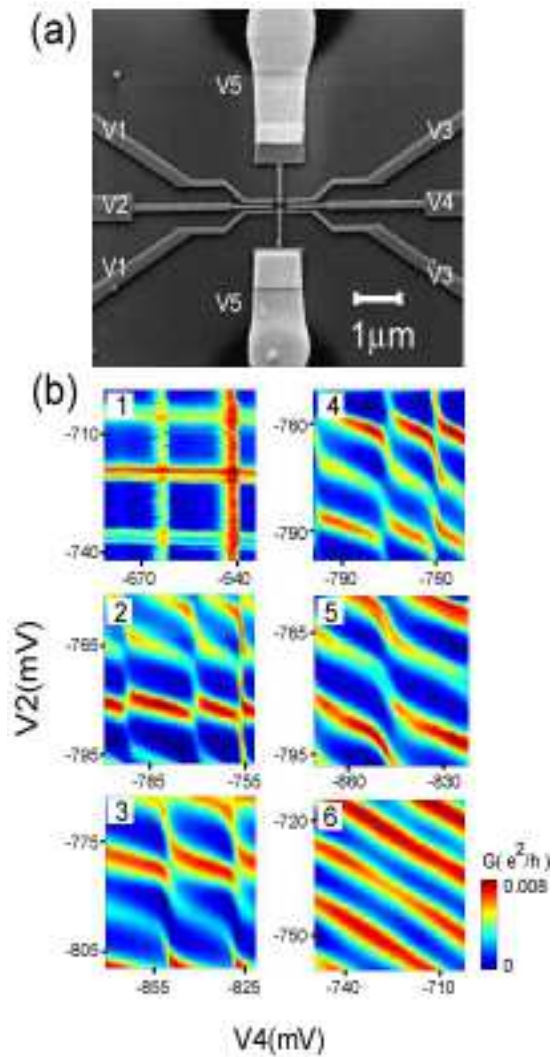}
\caption{\label{fig:epsart} Device and characterization in the 
CB regime: (a) Scanning electron micrograph of the device. 
(b) Logarithm of double dot conductance as a function 
of gate voltages V2 and V4 for weak lead-dot coupling. The color scale 
indicates the magnitude of conductance. The voltages in central pincher 
gate V5 are (1)-0.7477V (2)-0.6573V (3)-0.6534V (4)-0.6504V (5)-0.6494V 
(6)-0.5701V.}
\end{figure}
 
      Our device was fabricated on $GaAs/Al_xGa_{1-x}As$ heterostructure 
containing a two- dimensional (2D) electron gas 80nm below the surface, with 
electron density and mobility of $n=3.8\times 10^{11} cm^{-2}$ and 
$\mu = 9 \times 10^5 cm^2/Vs$, respectively.  The lithographic
size of each dot is $170nm \times 200nm$ (see Fig. 1(a)).  The eight
separate metallic gates
are configured and operated with five independently tunable gate voltages.
The dark regions surrounding (and underneath) gates 5 represent 120nm thick
over-exposed PMMA, which serve as spacer layers to decrease the local
capacitance thus preventing depletion, enabling each lead to simultaneously
connect to both dots.  The experiment carried out at a lattice-temperature of
30mK.  Standard, separate
characterization of each dot in the closed dot regime \cite{17} yielded a 
charging energy U of 2.517meV (2.95meV) for the left (right) dot, with 
corresponding dot-environment capacitance $C_\Sigma \approx 63.6af (54.3af)$, 
and level spacing $\Delta E \approx 219 \mu eV (308 \mu eV)$.  Modeling the 
dot as a metal disk embedded in a dielectric resulted in disk of radius 
$r_e \approx 70nm (60nm)$ with 58 (43) electrons.  To characterize the 
DQD and demonstrate the full tunability of our device, in Fig. 1(b) we 
show the Coulomb blockade (CB) charging diagram of the conductance versus 
plunger gates V2 and V4 \cite{18} as t was increased, for weak coupling to 
the leads where Kondo correlation is unimportant. The central pincher gate 
V5 controlled t where a reduction of the gate voltage decreased t. At weak 
coupling (t small), the electrons separately tunnel through the two nearly 
independent dots forming grid like pattern, yielding rectangular domains in 
Fig. 1(b)(1).  With increasing t charge quantization in individual dots is 
gradually lost as the domain vertices separated and the rectangles deformed 
into rounded hexagon. At large t when separate charge quantization is fully 
relaxed, the two dots merge into one and the domain boundaries become straight 
lines (Fig. 1(b)(6)). The evolution of the conductance pattern demonstrates 
the tunability of our DQD from ionic to covalent bonding states and bodes 
well for quantum computation applications \cite{19}.
         
          The parameters $\Gamma$ and t govern the delicate DQD Kondo physics.  
To obtain the necessary large t, the center pincher gate V5 was set so that 
the zigzag pattern in the charging diagram is barely visible (Fig. 2(a)), 
ensuring that the Kondo valleys can be located. To accomplish the formation 
of Kondo states in both dots, pincher gates V1 and V3 were tuned to give a 
sizable $\Gamma \approx \Delta E$ to ensure strong Kondo correlation.  An 
estimate for t is based on the fact that the charging diagram indicates a 
configuration close to the limit of a merged, single large dot (Fig. 2(a), 
so that the level broadening $\pi|t|^2/\Delta E$ should be comparable to the 
level spacing $\Delta E$, yielding $t\approx 150 \mu eV$, while $\Gamma$ is 
deduced to be $\approx 150 \mu eV$ from the half-width of the Coulomb blockade 
peaks.   
          
\begin{figure}
\includegraphics{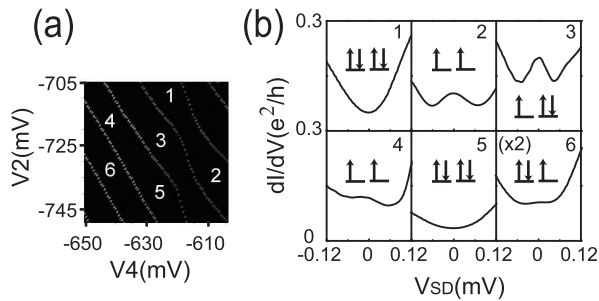}

\caption{\label{fig:epsart} Device characterization in the Kondo regime: 
(a) Charging diagram for the third cool-down depicted in a gray-scale plot 
of the conductance crest as a function of gate voltages V2 and V4 
(V5= -0.5965V). (b) Differential conductance traces from valleys 1 to 6 
in Fig. 2(a). 
The insets indicate the 
spin configuration of the uppermost, occupied electronic levels on each 
dot. Note that a single upward-pointing arrow only denotes an unpaired 
electron, and is not intended to represent the actual direction of spin 
alignment.}

\end{figure}        
          
        Care is required when changing t.  The mutual capacitive coupling 
between the gates and dots gives rise to a complex capacitance matrix.  
Operating V5 to tune t simultaneously affected the charge on the dots and 
other gates. Furthermore, slight residual asymmetry in the coupling to the 
leads caused the maximum of the Kondo zero-bias anomaly (ZBA) in the dI/dV 
to occasionally occur at nonzero voltages $V_{SD}\neq 0$ (termed the 
anomalous Kondo effect \cite{20}). By adjusting a combination of the 
remaining gates we tuned the ZBA to be nearly symmetrical about zero-bias. 
For practical purposes we relied on plunger-gates V2 or V4, which were found 
to follow V5 in an approximately linear manner.  This procedure can be 
thought as experimentally diagonalizing the capacitance matrix.  The 
desired configuration of an unpaired excess-spin on each dot could be 
maintained within a tuning range of $\approx 4 \sim 7mV$ in V5 without 
causing a sudden change in the charge configuration.  Note that V5 is set 
typically $\approx 70mV$ above pinch-off.  Therefore the relative tuning 
range in this Kondo regime is roughly 6-10\% of closure and even smaller 
for the corresponding fractional change in t, since tunnel-coupling becomes 
exponentially suppressed in the small value limit. 
          
         Three distinct spin configurations may appear assuming even-odd 
electron filling and focusing on the topmost states in each dot (Fig. 2(b)). 
Our investigation was carried out in such regimes. Non even-odd behavior 
was also observable, but will not be discussed \cite{21}.  The Kondo valleys 
and spin states in each dot were identified by measuring the differential 
conductance, dI/dV, versus source-drain bias, $V_{SD}$, at an electronic 
temperature of $ \sim 40mK$.  For regions 1-6 in Fig. 2(a), we observed 
the expected appearance and disappearance of Kondo resonance peaks near 
zero-bias shown in Fig. 2(b). 
     
\begin{figure}
\includegraphics{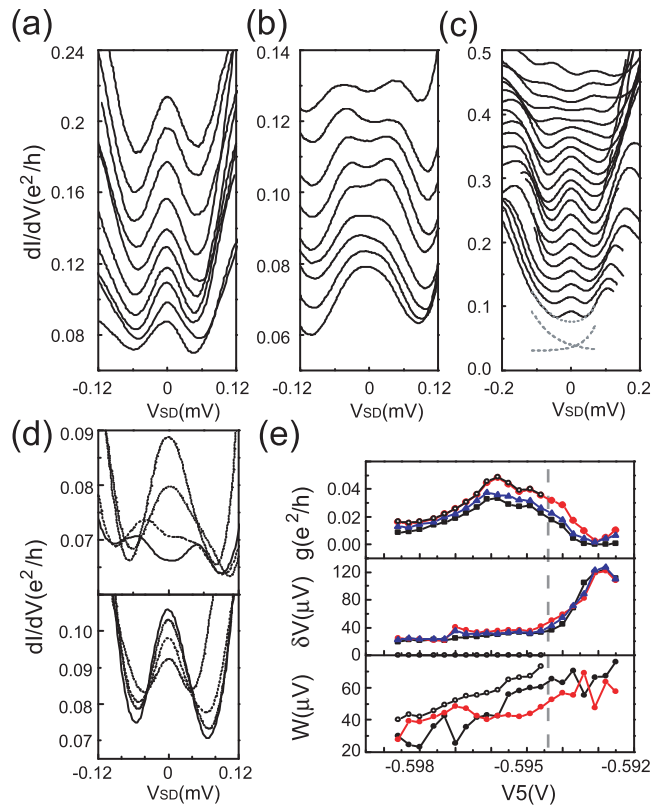}
\caption{\label{fig:epsart} The differential conductance, dI/dV, versus 
$V_{SD}$ for different inter-dot coupling strength tuned via V5 gate. 
(a) From top to bottom: V5= -0.6155, -0.616, -0.6165, -0.617, -0.6175, 
-0.618, -0.6185, -0.619, -0.6195, -0.62V, second cool-down. (b) From top 
to bottom: V5= -0.5965, -0.5970, -0.5975, -0.5978, -0.5982, -0.5987, -0.5990, 
-0.5992, -0.5995V, first cool-down, offset by 0.005e2/h for each trace. 
(c) From top to bottom: V5=-0.5925, -0.5928, -0.5931, -0.5934, -0.5937, 
-0.5940, -0.5943, -0.5946, -0.5949, -0.5952, -0.5955, -0.5958, -0.5961, 
-0.5964, -0.5967, -0.5970, -0.5973, -0.5976, -0.5980, -0.5983, -0.5986V, 
third cool-down as measured in Kondo valley 2 of Fig. 2(b), offset by 
$0.02e^2/h$ for each trace.  The traces vary in different cool-downs
due to the rearrangement in the occupation of charged in defects and traps.
(d) Selected data from (c) without offset.
Bottom half shows first quantum state regime (from top to bottom: 
V5=-0.5961, -0.5964, -0.5970, -0.5986). The upper half shows the second 
regime (from top to bottom: V5=-0.5946, -0.5940, -0.5937, -0.5934). (e) ZBA peak amplitude $g=dI/dV|_{V=0}$, peak splitting 
$\delta V$ and width W extracted by first subtracting three types of 
background signals followed by a two-Gaussian fit. The fit and data
are virtually indistinguishable.  The simulated
backgrounds are \textcolor{red}{$\bullet$}; two exponential functions 
+ constant (e.g. marked by the two gray dash lines for the bottom trace 
in (c) and combined functions shown as gray dot line); 
\textcolor{blue}{$\blacktriangle$} two Boltzmann and $\blacksquare$ 
linear function. The bottom figure shows the two peak widths obtained 
after the subtraction of a two exponential  background. 
Results for a single Gaussian fit (V5 $<$ -0.594V only) are indicated by 
open-circles ($\circ$).  Typically, the fitting error is smaller than 
the symbol. 
}
\end{figure}

            When t was tuned two distinct regimes of behavior were evident 
in dI/dV (see Fig. 3(a)-(d)).  The main features in the first regime 
(see Fig. 3(a),(d) bottom half) were the clear presence of a single peak 
in the ZBA, an increasing peak width and an increasing linear conductance 
$g=dI/dV|_{V=0}$ with increasing t.  In the second regime where t was 
increased further (see Fig. 3(b), (d) upper half), the single ZBA peak 
developed into two peaks. In some Kondo valleys, both types of behavior 
were observed as shown in Fig. 3(c),(d), in which the transition is seen 
to take place in a continuous manner. In the transition region, the ZBA 
peak broadened and its zero-bias value, g, approached a maximum, became flat 
before dropping as the ZBA peak gradually split into two. The suppression 
of g on the double-peaked side can be attributed to the emergence of 
spin-singlet correlation between the two dot spins \cite{1,2,3,4,5,6,7}.  
The maximum g is about $\sim  0.1e^2/h$, reduced below the theoretical 
unitary limit of $4e^2/h$. This reduction can occur when the energy levels 
in the two dots differ and the dot-lead coupling is asymmetric \cite{3,22}. 
            
         Analyzing the transition with detailed curve fitting we quantify 
the amplitude, splitting and width of the ZBA peak(s) at each V5 gate voltage 
within well-defined bounds. In the absence of precise theoretical functional 
forms either for the conductance background or the ZBA peak itself, we first 
systematically subtracted three sensible backgrounds.  After subtraction of 
each background we fitted the ZBA using three (combined) peak shapes: one 
(or two) Gaussian, Lorentzian and the Breit-Wigner resonance form. The 
resultant parameters invariably exhibited similar trends as shown in 
Fig. 3(e). The double-peak feature visibly disappeared at $V5\sim -0.594V$ 
as indicated by the dash line, and roughly coincides with the g maximum 
position.  Note that the peak splitting,$ \delta $, dramatically reduces 
from a maximum of $ \sim 120 \mu eV$ to $\sim  0$ when V5 is slightly 
reduced from -0.5925 V to $\sim$ -0.595 and correspondingly t is reduced by 
less than 4\% from an initial value $t\approx 150 \mu eV$.

      In addition to differences in the peak shape the 
temperature dependence of dI/dV was also distinct in the two regimes 
(see Fig. 4). In the single peak regime, the zero-bias dI/dV, g(T), 
decreased logarithmically with T (Fig. 4(a)). In contrast, when double 
peaks appeared, g(T) exhibited a non-monotonic behavior (Fig. 4(b)),  
where with increasing T g(T) slightly increased initially, then slowly 
decreased before increasing again when T exceeded $T_K$.  Together, these 
evidences point to qualitatively different phases in the two regimes and 
the existence of a quantum transition between them.
 
\begin{figure}
\vspace*{0.5cm}
\includegraphics{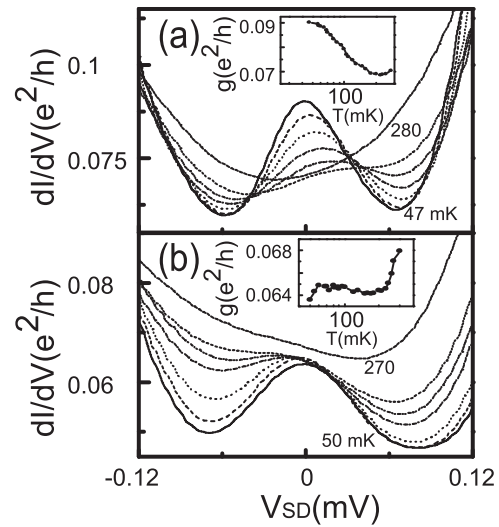}
\caption{\label{fig:epsart} Temperature dependence of the differential 
conductance dI/dV versus $V_{SD}$ in the Kondo valley of Fig. 3(c): 
(a) First quantum state regime: from top to bottom T=47, 70, 95, 110, 150, 180, 
220, 280 mK, at V5=-0.5953V. (b) Second quantum state regime: 
from bottom to top T= 50, 80, 100, 160, 180,210, 250, 270 mK, at 
V5=-0.5928V. Insets show the distinctly different temperature dependences 
of the zero-bias linear conductance, $g(T)=dI/dV|_{V=0}$, in the two regimes.}

\end{figure}   
         
         To date no theoretical work has addressed the parallel-coupled DQD 
with inter-dot tunnel coupling.  Nevertheless, because the ZBA occurs 
under slightly non-equilibrium conditions it is likely that we may identify 
the observed transition with the quantum critical phenomenon discussed 
in the two-impurity Kondo problem for the series geometry based on the 
following evidence:  (a) a continuous evolution from the single- to double 
peaked behavior, (b) a maximum in g, the zero-bias dI/dV, near the 
transition point, (c) different behaviors in the temperature dependence 
of g, and (d) a strong renormalization of the peak splitting, $ \delta $, 
compared to the estimated t or AF coupling, J, close to the transition.  
These features are all in qualitative agreement with predictions for a 
series-coupled DQD \cite{1,2,3,4,5,6,7,22}, although ideally theory predicts 
a maximum g reaching the unitarity limit $4e^2/h$.  Semi-quantitively, it 
is informative to roughly estimate key parameters and compared these to 
observed splitting, $ \delta \leq 120 \mu eV$.   Within the theoretical 
scenarios, $\delta$ must be compared to $4t \approx 600 \mu eV$ or 
$2J = 8t^2/U \approx 180 \mu eV$.  (Note that in the open dot regime U 
is expected to be reduced from its closed dot value by roughly 1/3 \cite{11}, 
yielding $U \approx 1meV$.)   The reduction of $\delta$ compared to 4t 
and 2J are in agreement with theory and lends further credibility to our 
identification of the quantum transition. 
            
{\it Acknowledgements}  We thank H. Baranger, K. Matveev, 
S. Khlebnikov, N. Giordano, N. Windgreen, 
D. Cox, A. Schiller, and H. Nakanishi for discussions, and F. Altomare, 
L.C. Tung, V. Gusiatnikov, and H. Jeong for assistance in the experiment.  
Work supported by NSF grants DMR-9801760 and DMR-0135931.


\end{document}